# Quasi-bound flat bands in the continuum

Haoyu Qin[1,2,*], Weixuan Zhang[1,2,*,#], Shaohu Chen[3,*], Huizhen Zhang[1,2], Ruhao Pan[4], Junjie Li[4], Lei Shi[3,+], Jian Zi[3], and Xiangdong Zhang[1,2,$]

[1]Key Laboratory of Advanced Optoelectronic Quantum Architecture and Measurements of Ministry of Education, School of Physics, Beijing Institute of Technology, Beijing 100081, China.

[2]Beijing Key Laboratory of Nanophotonics and Ultrafine Optoelectronic Systems, School of Physics, Beijing Institute of Technology, Beijing 100081, China.

[3]Key Laboratory of Micro- and Nano-Photonic Structures (Ministry of Education), Department of Physics, Fudan University, Shanghai 200433, China.

[4]Beijing National Laboratory for Condensed Matter Physics, Institute of Physics, Chinese Academy of Sciences, Beijing 100190, China.

*These authors contributed equally to this work.

[$+#]To whom correspondence should be addressed. E-mail: zhangxd@bit.edu.cn; lshi@fudan.edu.cn; zhangwx@bit.edu.cn

**Bound states in the continuum (BICs) are widely known spatially localized states experimentally implemented as quasi-BICs. Although they emerged as a promising solution for achieving high-quality resonances in photonic structures, quasi-BICs are confined to a very narrow range in k-space and are highly sensitive to disorder. Here, we introduce quasi-bound flat bands in the continuum (quasi-BFICs) — a class of optical states where Bloch modes are found within a photonic flat band, leading to a quasi-BIC behaviour at every k-point above the light line. We analytically and numerically demonstrate the origin of quasi-BFICs from the disorder-induced band folding, mode localization and multiple topological charges in k-space, and identify the optimal strength of structural disorder to maximise their generation probability. Angle-resolved transmission and Q-factor measurements confirm the existence of quasi-BFICs, opening new avenues for designing devices with high quality factor and wide-angle response, presenting a counterintuitive strategy that leverages disorder to enhance optical performance.**

Trapping electromagnetic waves in micro- and nanostructures with high quality factors (Q-factors) is critical for enhancing light-matter interactions and advancing novel photonic devices. Over the past decade, bound states in the continuum (BICs) [1-8]—a unique class of localized states whose energies reside within the continuum of radiating modes—have garnered significant attention as a promising approach to the design of photonic structures with exceptionally high Q-factors [2-52], enabling various innovations such as BIC lasers [9-13], ultra-sensitive sensors [14-17], and more [18]. In 2013, Hsu et al. made a groundbreaking experimental realization of photonic BICs in two-dimensional (2D) photonic crystal (PhC) slabs [2], demonstrating both symmetry-protected and accidental BICs at distinct $k$ points with theoretically infinite Q-factors. However, despite these progress, several issues remain in translating BICs to many practical applications. First, quasi-BICs with high theoretical Q-factors are typically confined to a narrow range in momentum space. This means that efficiently exciting quasi-BICs requires a precise excitation angle, as even slight deviations can lead to a sharp drop in the Q-factor. Second, quasi-BICs are highly dispersive in PhCs, limiting their performance under the wide-angle illumination with a fixed frequency. Lastly, the Q-factors of experimentally realized BICs are constrained by fabrication imperfections, such as surface roughness and structural disorder. These imperfections introduce additional radiation losses, reducing the Q-factors of quasi-BICs in practical applications.

To address these long-standing issues, several novel types of optical BICs have been proposed and realized. For example, a merged BIC [20-22], in which multiple topological charges [19] are tuned to lie very close to each other, can broaden the momentum range of quasi-BICs with high Q-factors and suppress out-of-plane-scattering losses caused by structural disorder. In addition, by merging a symmetry-protected BIC with two Friedrich-Wintgen quasi-BICs, researchers have developed a super BIC, which features an enhanced Q-factor and near-zero group velocity across an extended region in the Brillouin zone [23, 24]. Although merging multiple BICs or quasi-BICs can locally broaden the $k$-space range of high-Q resonances and reduce band dispersion, their performances remain limited in the entire $k$-space. More recently, a moiré BIC, protected by multiple topological charges locating at different orders of diffraction channels, has been theoretically proposed and experimentally demonstrated [25]. This approach further extends the momentum range of quasi-BICs while improving band flatness. However, the Bloch modes that are distant from moiré BICs still exhibit low Q-factors. Building on these advances and addressing the remaining limitations, a promising direction is the construction of a photonic flat band, where the Bloch mode at each $k$-point above the light line behaves as a quasi-BIC. We refer to this unique optical state as a quasi-bound flat band in the continuum (quasi-BFIC), characterized by its capacity to sustain high-Q resonances across the full range of excitation angles (Figure 1a). In contrast, current quasi-BICs are confined to a very narrow angular range, where the radiation loss of nearby guided resonances increases rapidly as they move away from the BIC (Figure 1b). This distinctive feature of quasi-BFICs is expected to open new opportunities for enhancing light-matter interactions under wide-angle illumination, thereby broadening the scope of BIC-based applications. A crucial question remains: how can such an innovative optical state be effectively realized?

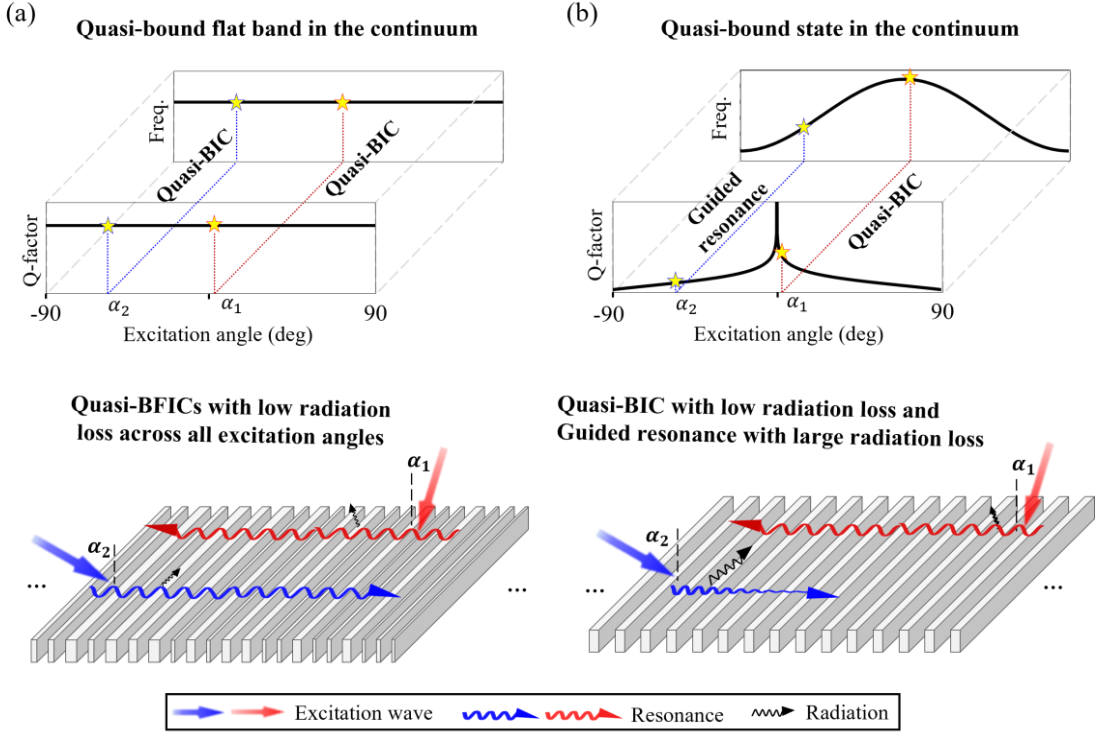

**Figure 1. Illustration of the superior robustness of quasi-BFICs to the excitation angle compared to quasi-BICs.** (a). High-Q resonances of quasi-BFICs extend across the entire *k*-space, showing the strong robustness to the excitation angle. (b). Conventional quasi-BICs are confined within a narrow range in momentum space. Even a slight deviation of the excitation angle can lead to a significant drop in the Q-factor and an increase in the far-field radiation loss.

On the other hand, the field of photonic systems has witnessed a remarkable surge of interest in exploring disorder effects, with numerous studies demonstrating that disorder can both induce localization phenomena (such as random laser emission [53] and Anderson localization of light [54]) and enhance wave transport [55, 56]. While, concerning BICs in PhC slabs, all existing experimental observations have unveiled substantial scattering of Bloch waves due to disorder [2, 20], which compromises the performance of BICs. In this work, we challenge the detrimental view of disorder on BICs and reveal that quasi-BFICs can be randomly generated in PhC slabs through mirror-symmetric disorder within each supercell. Using both analytical and numerical methods, we demonstrate that random quasi-BFICs arise from the combined effects of band folding, mode localization and multiple topological charges induced by disorder. Additionally, we identify an optimal strength of structural disorder that maximizes the probability of generating random quasi-BFICs, which results from the balance between disorder-induced modal localization, band flatness, and the radiative Fourier coefficients of the localized eigenmode. In experiments, we fabricate a randomly created PhC slabs and observe quasi-BFICs through both angle-resolved transmission spectra and iso-frequency contour image of the scattering light. Our findings open new avenues for the design of highly efficient optical devices with wide-angle responses, while also suggesting a counterintuitive strategy that leverages disorder to improve the optical functionality.

**The theoretical framework for the quasi-bound flat bands in the continuum.** We start with a one-dimensional (1D) silicon PhC slab, as shown in Figure 2a, with the system parameters detailed in the figure caption. Figure 2b presents the calculated band structure of TE-like modes, with the colormap representing the corresponding Q-factors. We find that an isolated BIC with infinite Q-factor is identified at the center of momentum space, while other Bloch modes located within the light cone (the white region), experience a rapid decrease in Q-factor as their $k$-vectors deviate from the BIC. In addition, Bloch modes below the light line are non-radiative, exhibiting infinite Q-factors. Now, we construct a supercell with thirteen fundamental units ($A=13a_0$) by adding mirror-symmetric disorder. Then, the supercell is repeated in periodic direction to construct an infinite photonic crystal slab. Specifically, as depicted in Figure 2c, the widths of silicon pillars within each supercell are randomly assigned within the range of [$W$-$\delta$, $W$+$\delta$], where $W$ is the original width and $\delta/W$ characterizes disorder strength. We emphasize that that the mirror symmetry of each supercell is essential for constructing the quasi-BFIC, which is fully discussed later. To maintain the mirror-symmetry of each supercell, pairs of silicon pillars positioned symmetrically about the mirror plane share the same width. We mention that the proposed structure is different from the chirped system with gradually varying structural modulations and exhibit local rather than global periodicity [57-59]. The disorder adopted in the random design breaks the local periodicity but preserves the global periodicity, as the supercell is periodically repeated to form the final structure.

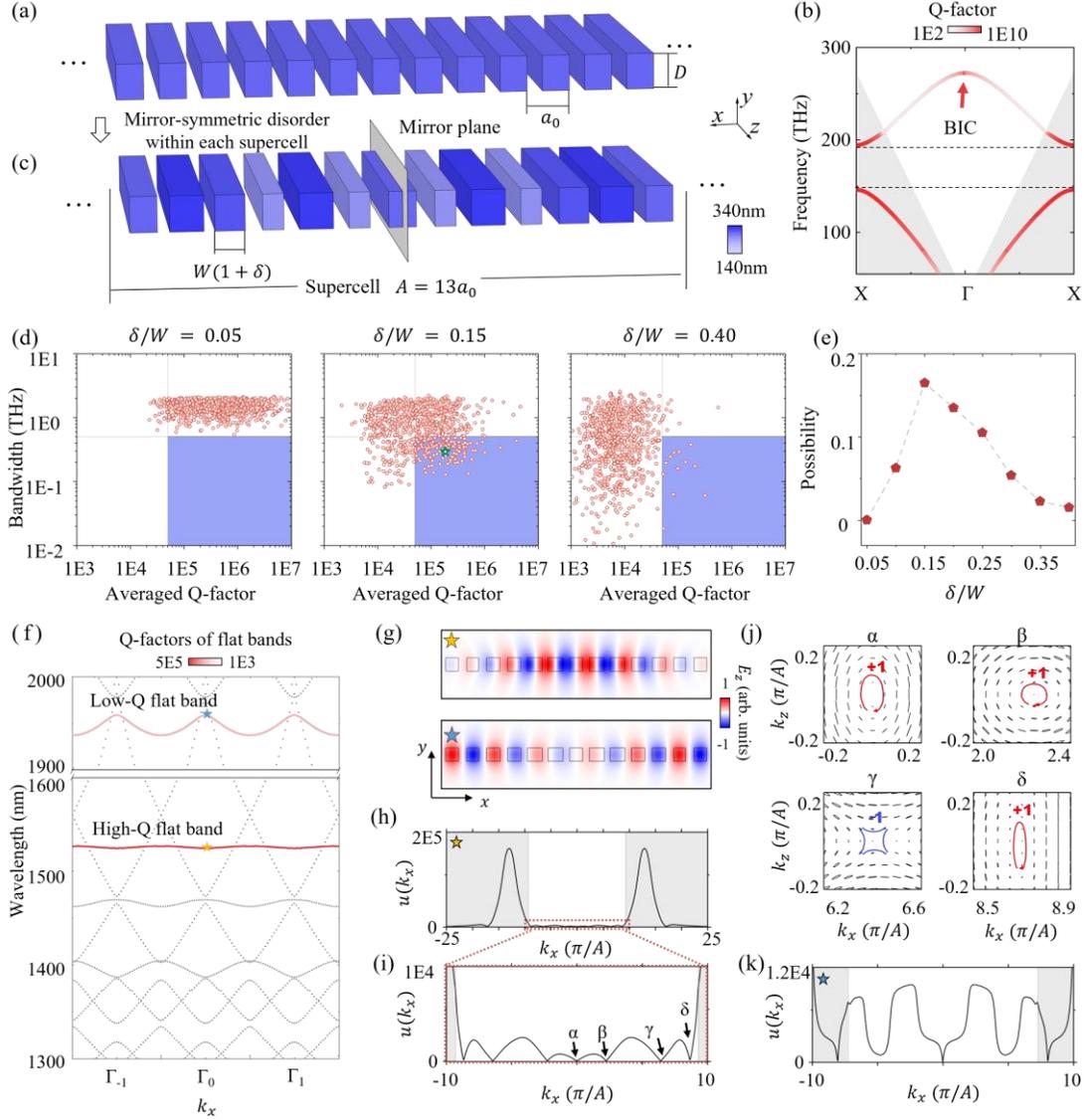

**Figure 2. Theoretical results of quasi-bound flat bands in the continuum.** (a) Schematic diagram of the 1D silicon PhC slab. The lattice constant, slab thickness and silicon width are denoted as $a_0$=380 nm, and $D$=250 nm and $W$=243 nm, respectively. The refractive indexes of background and silicon grating are $n_b$=1.45 and $n_{Si}$=3.33. (b) The calculated TE-like band structure of a 1D silicon PhC slab, with the colormap representing the Q-factors of Bloch modes. (c) Schematic diagram of a supercell in the 1D silicon PhC slab, with mirror-symmetric disorder on $W$ within each supercell. Each silicon grating is centered within a fundamental unit, and the widths of the pillars are randomly assigned in the range [$W$-$\delta$, $W$+$\delta$], where $\delta/W$ characterizes the disorder strength. The color of each pillar indicates its width after introducing disorder. (d) Statistical distributions of bandwidths and averaged Q-factors of a targeted TE-like band for perturbed structures with $\delta/W$ equaling to 0.05, 0.15, and 0.4. Structures with high-Q flat bands are enclosed in the blue shaded areas. (e) Probability of generating random structures with high-Q flat bands (dots in the blue blocks) as a function of disorder strength. (f) Calculated band structure of a randomly generated PhC slab, as marked by the green star in Fig. 2d. Two flat bands are highlighted, and the colormap represents their Q-factors. Other Bloch modes are shown as gray lines. $\Gamma_n$ with $n = 0, \pm 1$ represent the $\Gamma$-point in Brillouin zones of different orders. (g) Upper: Eigen-field profiles at the $\Gamma$ point for quasi-BFIC; lower: Eigen-field profiles at the $\Gamma$ point for low-Q flat band. (h) Calculated amplitude distributions of Fourier coefficients for flat-band eigen-fields in $k$-space $u(k_x)$ for the high-Q flat band at 1520 nm. (i) The enlarged distribution of $u(k_x)$ for the high-Q flat band

at 1520 nm. (j) Multiple topological charges related to zero points of $u(k_x)$, which are marked by characters of α, β, γ, and δ in (i). (k) The enlarged distribution of $u(k_x)$ for the low-Q flat band at 1935 nm.

To assess the impact of mirror-symmetric disorder with different strengths, we randomly generate 800 different structures at each value of $\delta/W$, ranging from 0.05 to 0.4. We then calculate the bandwidth and averaged Q-factor of the first TE-like band above the low-frequency bandgap (marked by the dash lines in Fig. 2b) of these structures. Notably, our considered band is isolated from other TE-like bands due to band folding and coupling induced by the structural disorder within each supercell (Supplementary Note 1). The bandwidth is defined as the difference between the maximum and minimum eigen-frequencies of the targeted band (Supplementary Note 2). And, the averaged Q-factor is calculated as the mean of the Q-factors for all Bloch modes within the band. We present three typical statistical distributions with disorder strengths being $\delta/W$ =0.05, 0.15 and 0.4 (Figure 2d). When $\delta/W$ = 0.05, most randomly created structures exhibit high averaged Q-factors of the targeted band, but also relatively large band dispersions. As $\delta/W$ increases to 0.15, contrary to conventional expectations that disorder would reduce the Q-factors, we observe that some randomly generated structures can exhibit both high averaged Q-factors and narrow bandwidths, as indicated by discrete points within the blue region. However, when the disorder strength is further increased to $\delta/W$=0.4, the averaged Q-factors of all generated structures drop significantly. The probability of generating PhC slabs with high Q-factor flat bands (dots in the blue region) reaches maximum at $\delta/W$=0.15 (Figure 2e).

Because when δ/W is relatively small, the eigenmodes of these randomly generated structures always exhibit weak localization within each supercell, accompanied by significant band dispersion. Conversely, if the disorder strength is too large, the strongly localized eigenmodes within each supercell can induce the substantial far-field radiation. Therefore, the optimal strength of structural disorder is determined by balancing the disorder-induced modal localization and band flatness, as well as the amplitude of Fourier coefficients above the light line (the far-field leakage) of the localized eigenmodes (see details below).

To further illustrate the disorder-induced high-Q flat bands, we calculate the band structure of a randomly generated PhC slab with $\delta/W$ = 0.15 (marked by the green star in Fig. 2d) sustaining the high-Q flat band (Figure 2f). Notably, two flat bands are highlighted in the eigen-spectrum, with the colormap representing their Q-factors. Other dispersive Bloch modes are plotted by gray lines. Remarkably, flat band at 1520 nm exhibit exceptionally high Q-factors, reaching up to 200000 throughout the entire *k*-space, respectively. We emphasize that such angle-insensitive Q-factors can substantially enhance robustness against scattering losses from fabrication defects (Supplementary Note 3). Another flat band at 1935 nm has the relatively low Q-factor (about 8000). Both flat-band eigenmodes exhibit strong localization within the supercell (Figures 2g). These spatially localized eigenstates induce the formation of the flat-band dispersions.

To analyze the origin for the high Q-factors of flat bands, we calculate the amplitude distribution of Fourier coefficients for flat-band eigen-fields in momentum space [19]. It is defined as

$$u_n(\tilde{k}_x) = |\frac{1}{A}\int_A \mathbf{E}(\tilde{k}_x, x, y)e^{ik_x x}dx| \qquad (1)$$

with $k_x = \tilde{k}_x + n\frac{2\pi}{A}$ ($n = 0, \pm1, ...$). $\mathbf{E}(\tilde{k}_x, x, y)$ represents the flat-band eigenmode at the Bloch wavevector $\tilde{k}_x \in [-\frac{\pi}{A}, \frac{\pi}{A}]$ within a supercell out of the slab. For simplicity, we denote the set of $u_n(\tilde{k}_x)$ for all permissible values of *n* as $u(k_x)$. Here, we set the integral plane of Eq. (1) in the near-field region, making both radiative and non-radiative components of flat-band eigenfields in *k*-space can be obtained. If the integral plane is set in the far-field region, Eq. (1) only quantifies the far-field radiation amplitude [19] (Supplementary Note 4). Details on the calculation of $u(k_x)$ using the finite element method is presented in the Methods section. It is shown that, for the high-Q flat band at 1520 nm the dominate components of $u(k_x)$ are concentrated in the non-radiative region, with extremely low values of $u(k_x)$ appearing in the radiative domain (Figure 2h). This suggests that only a small portion of the Fourier coefficients for the flat-band eigenmode is able to radiate away from the PhC slab.

Additionally, the suppression of radiative components above the light cone is further enhanced by the presence of multiple zero points of $u(k_x)$. When we magnify the view of $u(k_x)$ in the radiative area (Figure 2i), multiple zero-amplitude points (labeled as α, β, γ, and δ) are observed at discrete *k* points. These zero points are characterized by ill-defined far-field polarization, corresponding to the polarization singularities with topological charges equaling to $\pm1$ (Figure 2j). The robustness of the topological charges is proved by their evolution under continuous parameter variation (Supplementary Note 5). We note that, beyond preventing leakage at specific wavevectors, topological charges also suppress the radiation of Bloch modes at nearby wavevectors, which explains the low values of $u(k_x)$ in the radiative region. In this case, protected by the existence of multiple zero points of $u(k_x)$, each Bloch mode within the high-Q flat band can be regarded as a quasi-BIC. Consequently, this high-Q flat band behaves as a quasi-BFIC. For comparison, we further calculate the distribution of $u(k_x)$ for the low-Q flat band at 1935 nm, as shown in Figure 2k with enlarged plot within the radiative region. Unlike the high-Q flat band, there is only one topological charge within the radiative region, resulting in the significantly increased values of $u(k_x)$ in this area. Consequently, this flat band possesses a large radiation loss and exhibits a much lower averaged Q-factor.

To further elucidate the above phenomena, we truncate the near-field distribution at $y_0$=100 nm for the eigenmode presented in the upper plot of Fig. 2g, as shown by black dots in Figure 3a. This flat-band eigen-field exhibits an exponentially localized spatial profile combined with an oscillating phase within a supercell, taking the form of $\mathbf{E}(x, y_0) = E_0 e^{-\alpha x^2} \sin(k_s x)$. The mirror-symmetric disorder ensures that the eigen-field is an even or odd function with respective to the mirror plane of the supercell. Here, the fitting parameter $\alpha$ determines the strength of the spatial localization and $k_s$ describes the wavevector of the oscillating phase. The red line in Figure 3a shows the fitted profile of this flat-band eigen-field, with fitting parameters being $\alpha = 4.40 \times 10^{11}\ [1/m^2]$ and $k_s = 8.44 \times 10^6\ [1/m]$. It is shown that the fitted eigenmode reveals an excellent agreement with the numerical counterpart.

Next, we demonstrate how these two fitting parameters play key roles in the generation of quasi-BFICs. It is important to note that the spatial profile of flat-band eigenfields are nearly independent of the Bloch vector due to the strong localization within each supercell. In this context, the amplitude distribution of Fourier coefficients for flat-band eigen-fields in momentum space can be re-expressed as

$$u_n(\tilde{k}_x) = |\frac{1}{A} \int_A \mathbf{E}(x, y_0) e^{i(\tilde{k}_x + n\frac{2\pi}{A})x} dx|, \tag{2}$$

with the out-of-slab eigen-field $\mathbf{E}(x, y_0)$ being independent of the Bloch vector. Figure 3b presents the calculated result of $u(k_x)$ using the analytical formula of $\mathbf{E}(x, y_0)$. We find that the analytical results closely align with the numerical findings in Figure 2h, showing two symmetrically positioned peaks in the non-radiative region. The wavevectors of two peaks match to the fitting parameter $\pm k_s$, which are located in the non-radiative domain below the light cone. Additionally, the widths of these two peaks are determined by the localization strength α, with larger values of α leading to broader peaks. For the flat band to exist, a sufficiently large α is required, ensuring that the eigenfields are strongly localized. However, if α is too large, two peaks in the non-radiative region broaden significantly, causing a substantial portion of $u(k_x)$ to spill into the radiative domain and thereby reducing the Q-factors (Supplementary Note 6). Thus, there exists an optimal range of the localization strength α, which is determined by the disorder strength (as the disorder strength increases, the value of α is also increased), allowing for the coexistence of the flat-band dispersion with localized eigen-modes and narrow peak-widths of $u(k_x)$ in the non-radiative domain. To further illustrate the radiative properties, we present an enlarged view of $u(k_x)$ around the center of *k*-space (Figure 3c). There are multiple equally spaced topological charges within the light cone and zero points of $u(k_x)$ in the non-radiative region near the light line, ensuring that the flat band possesses the low-valued radiation loss in the entire *k*-space. Interestingly, we find that the number of topological charges depends not only on the fitting parameters α, but also on the period *A* of the supercell. To verify this phenomenon, we calculate the variation in the number of topological charges (within the light cone) and the averaged values of $u(k_x)$ above the light line $<u(k_x)>_{|k_x|<k_b}$ ($k_b$ is the background wavevector) for the flat band as a function of the supercell period *A* (Figure 3d), with fitting parameters α and $k_s$ identical to those in Fig. 3a. The results indicate that the number of topological charges increases with the supercell period, accompanied by a significant reduction in the radiation loss. This suggests that expanding the supercell size could be a viable way to approaching ideal BFICs.

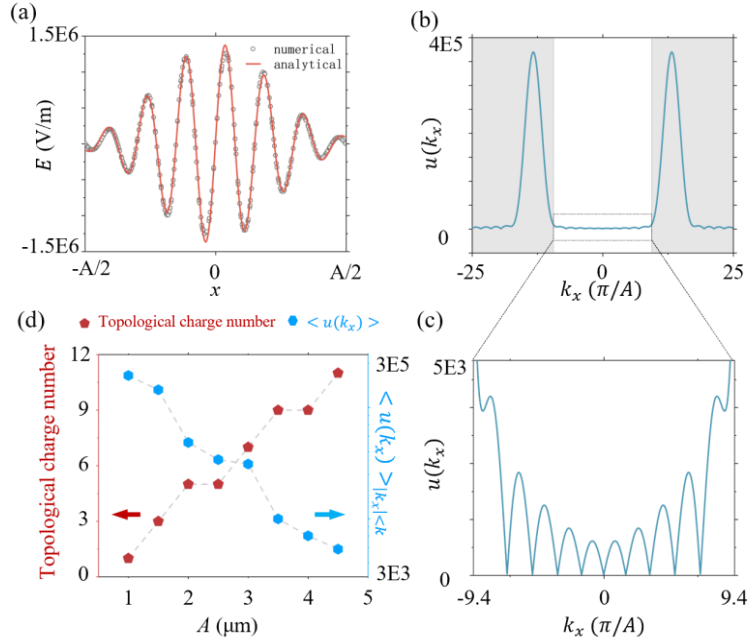

**Figure 3. The analytical explanation of quasi-bound flat bands in the continuum in randomly created photonic crystal slabs.** (a). The numerical (the black dots) and fitting results (the red line) of the near-field distribution of the flat-band eigenmode at *y*=100 nm, truncated from the upper plot of Fig. 2(g). Here, the fitting parameters $\alpha$ and $k_s$ of $\mathbf{E}(x) = E_0 e^{-\alpha x^2} \cos(k_s x)$ are equal to $\alpha = 4.40 \times 10^{11}$ [1/m$^2$] and $k_s = 8.44 \times 10^6$ [1/m], respectively. (b). The analytical result of the amplitude distribution of Fourier coefficients for the flat-band eigen-field in momentum space $u(k_x)$. (c). The enlarged view of the analytically calculated result of $u(k_x)$ above the light line. (d). Analytical results on the variation in the number of topological charges and the average values of

$u(k_x)$ above the light line for the flat band as a function of the supercell period $A$, with fitting parameters α and $k_s$ identical to those in (a).

Additionally, we note that besides the spatial profile described by $E_0 e^{-\alpha x^2} \sin(k_s x)$, other types of flat band eigenfields with varied distributions can also be randomly generated with mirror-symmetric disorder within each supercell. It is shown that these different spatial profiles of localized flat-band eigenmodes can also give rise to quasi-BFICs under suitable fitting parameters (Supplementary Note 7). Moreover, the importance of the mirror-symmetry on the emergence of quasi-BFICs is also theoretically verified by considering the flat-band eigen-field without any even or odd symmetry, where no topological charge is found above the light line (Supplementary Note 8).

**The experimental evidence for the quasi-bound flat bands in the continuum.** In this section, we provide the experimental results, including the angle-resolved transmission spectra, iso-frequency contour, and the direct Q-factor measurement of the quasi-BFIC resonances. To verify our findings, the photonic crystal samples are fabricated according to the theoretical parameters. Figure 4a and b shows a scanning electron microscope image of the fabricated sample, with an enlarged side view in Figure 4b. A 241 nm thick silicon layer is firstly deposited on a silica substrate. Then, the 1D PhC slab with mirror-symmetric disorder in each supercell is fabricated in the silicon layer using a combination of electron-beam lithography and reactive ion etching techniques (see Methods section for details). The fabricated PhC slab measures approximately 500×500 μm² and contains ~100 supercells with each supercell possessing a length of 4940 nm. Such sample size is sufficient to exhibit the high-Q properties associated with quasi-BFICs (Supplementary Note 9). To create an optically symmetric environment, the fabricated sample is covered with PMMA, which matches the refractive index of the silica substrate.

At first, we measure the angle-resolved transmission spectra using the optical setup shown in Figure 4c. Details are provided in Methods section. The result is shown in Figure 4d where the color map qualifies the strength of transmissivity. In Supplementary Note 10, the line spectra with specific incident angles are presented. The accessible ranges of the incident angle and wavelength in our applied polarization-resolved momentum-space imaging spectroscopy are confined within -40 to 40 degree and 1410 to 1580 nm, respectively. For comparison, we also numerically calculate the transmission spectra of this 1D PhC slab (Figure 4e). The scanning steps for the incident wavelength and wavevector are matched to their experimental resolution limits. We find that the measured dispersions of Bloch bands agree well with the simulation results, confirming the successful fabrication of the sample with mirror-symmetric disorder within each supercell. Notably, we find that for both the experimental and numerical transmissivities, the high-Q flat band vanish entirely across the full range of incident angles. As a matter of fact, this observation serves as indirect validation for the presence of quasi-BFICs. The disappearance of the entire band shares the same origin as the disappearance of small segment of the band observed in previous BIC studies [2, 24, 25, 29, 45].

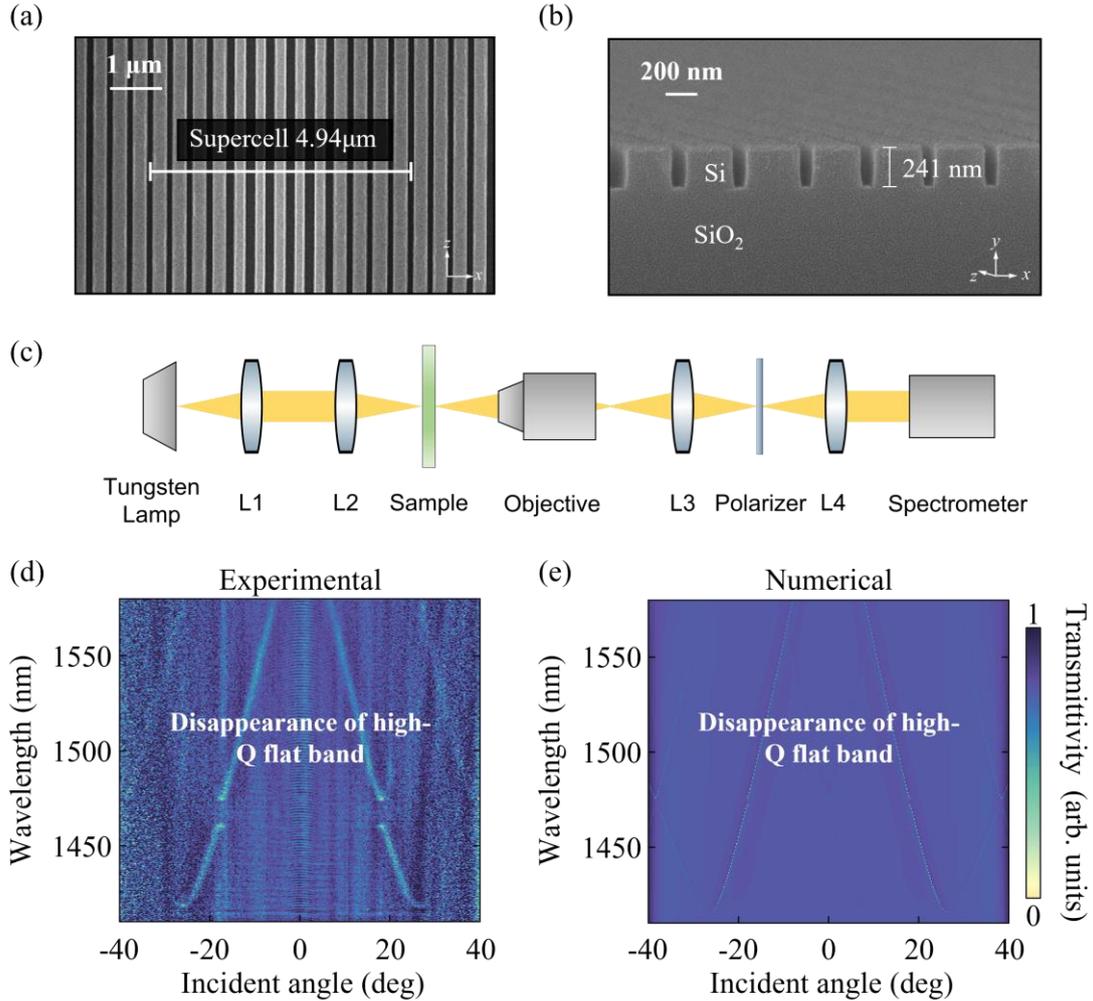

**Figure 4. The measurement of the angle-resolved transmittance spectra.** (a). Scanning electron microscope image of the fabricated 1D silicon PhC slab with mirror-symmetric disorder within each supercell. The supercell constant and slab thickness are equal to 4940 nm and 241 nm, respectively. In addition, the disorder strength applied to the width of this silicon grating is equal to $\delta/W=0.15$. (b). The enlarged side view of a supercell for the fabricated sample. (c) The optical setup of the angle-resolved momentum-space imaging spectroscopy. (d) and (e) The experimentally measured and numerically stimulated transmission spectra as functions of the incident wavelength and wavevector. The scanning steps for the incident wavelength and wavevector are matched to the experimental resolution limits for these quantities. High-Q flat bands are missing in these results.

To directly measure the Q-factors of quasi-BFICs, we employ an additional optical setup (Figure

5a), which enables band structure visualization through iso-frequency contour observation in momentum space. Details are provided in the Methods section. The experimental results of iso-frequency contours at 1520 nm, 1515 nm, and 1510 nm can be obtained through precise control of the incident point (the large bright spot in the upper right corner), where two distinct sets of iso-frequency contours are observed (Figure 5b-d). The flat-band iso-frequency contours appear as parallel lines that gradually separate with decreasing wavelength. The second set of contours corresponds to the dispersive TM-like mode. For reference, numerical simulations of the iso-frequency contours are displayed in Figure 5e-g, where gray and red lines represent the eigenmodes of the quasi-BFICs and dispersive bands, respectively. Excellent agreement between experimental results and theoretical predictions confirms the observation of flat bands.

The observation of flat bands enables direct extraction of the resonance Q-factor through scattering spectrum linewidth measurements. We select four resonances centered at 1515 nm (marked by colored symbols in Figure 5c), corresponding to incident angles ranging from 18.3 deg to 47.4 deg. By scanning the incident wavelength from 1513 nm to 1517 nm in 0.05 nm steps, narrow intensity peaks are observed with symmetric Lorentzian line shapes (Figure 5h-k), where Lorentzian fitting of these scattering spectra yields the Q-factor of the resonances. Remarkably, the Q-factor remains stable at approximately 8000 across the entire angular range. This contrasts sharply with traditional BICs, where the Q-factor drops rapidly to ~$1\times10^3$ with just 2 deg deviation from normal incidence [2]. Our proposed quasi-BFIC structure thus demonstrates exceptional angular insensitivity of the high-Q property.

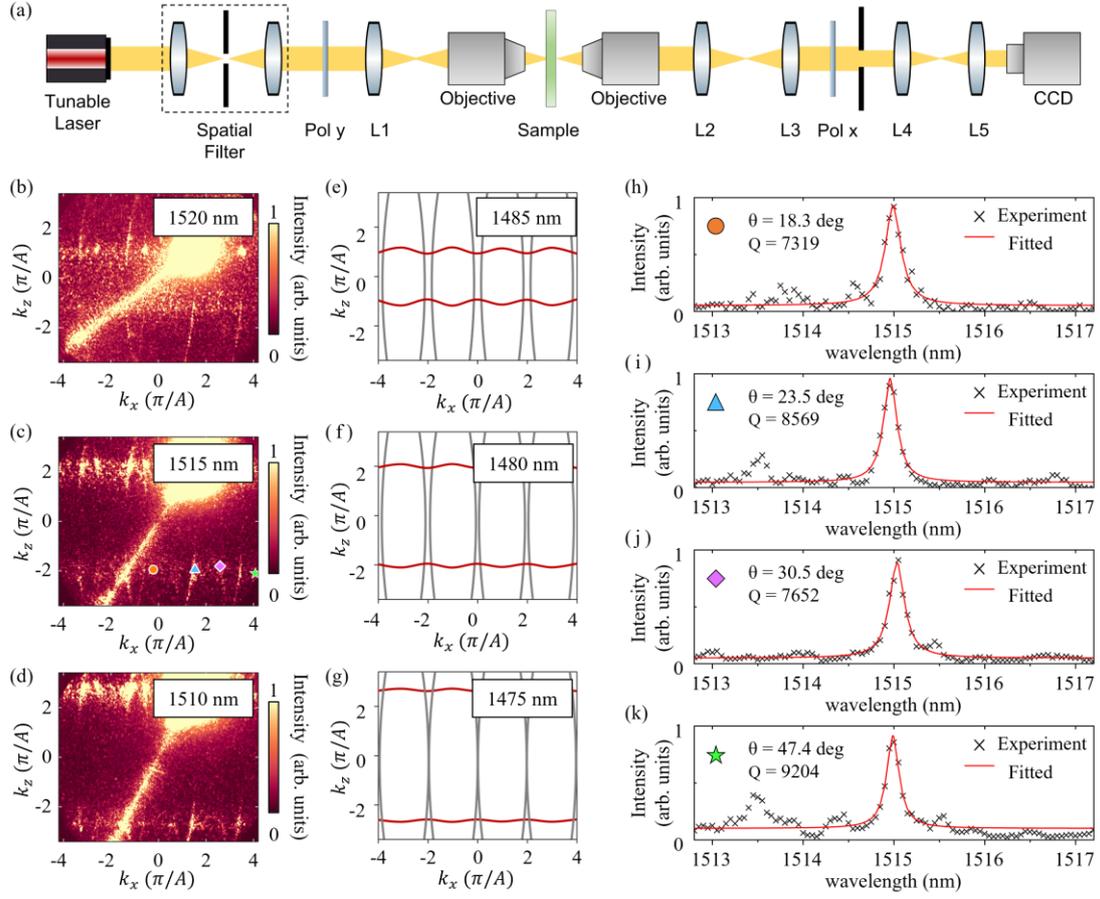

**Figure 5. Direct observation of quasi-BIC flat bands.** (a) The schematic of the isofrequency contour imaging system. (b)-(d) Experimental iso-frequency contours at 1520 nm, 1515 nm, and 1510 nm, respectively. The high-Q flat bands appear as diverging parallel lines with decreasing wavelength. (e)-(g) Corresponding numerical simulations at 1485 nm, 1480 nm, and 1475 nm, respectively, with gray and red lines representing the eigenmodes of high-Q flat bands and dispersive bands. (h)-(k) Experimental intensity spectra (dots) with Lorentzian fits (lines) for four resonances at the marked k-points in (b).

**Discussion and outlook.** In conclusion, we have introduced an approach to utilizing disorder in PhC slabs to create high-Q flat bands, which we term quasi-BFICs, with each Bloch state within this flat band exhibiting the quasi-BIC characteristics. Contrary to the conventional belief that disorder degrades the performance of photonic BICs, our study demonstrates that structural disorder can, in fact, enhance light-matter interactions. Additionally, we identify an optimal disorder strength that maximizes the probability of generating random quasi-BFICs. Through analytical and numerical analyses, we have shown that these random quasi-BFICs arise from the band folding, mode localization, and the emergence of multiple topological charges induced by disorder. Furthermore, we experimentally confirmed the existence of quasi-BFICs through both angle-resolved transmission spectroscopy and scattering iso-frequency contour imaging, demonstrating nearly uniform Q-factors of approximately 8000 across a wide angular range (up to 47.4 deg). These findings challenge the traditional understanding of disorder in photonic systems and open up exciting possibilities for designing advanced optical devices with enhanced functionality and responsiveness. Our work paves the way for the future advancements in photonic technologies, with the potential to revolutionize the design and applications of highly efficient, wide-angle response and disorder-tolerant optical systems based on BICs. Additionally, the concept of quasi-BFICs holds promise for extension to other classical wave and condensed matter systems, offering valuable insights for the development of novel functional devices and the exploration of flat-band quantum phenomena with many-body strong interactions [7, 60].

**Methods**

**Sample fabrication:** A 241 nm-thick silicon film was grown on a 500 μm-thick silica substrate using plasma-enhanced chemical vapor deposition (PECVD). This silicon layer exhibits the negligible intrinsic loss within the wavelength range from 1300 nm to 2000 nm, making it highly suitable for optical measurements. A 350 nm layer of ZEP520A electron-beam resist was then spin-coated onto the silicon film to serve as a photoresist. Following exposure and development, the resist patterns were defined and subsequently transferred to the silicon film using anisotropic etching with HBr plasma. The ZEP520A layer acted as an etching mask during this process, ensuring precision patterning on the silicon film.

**Angle-resolved transmittance spectrum measurement system:** Transmittance spectra were obtained using a home-built, polarization-resolved momentum-space imaging spectroscopy system integrated with a Nikon microscope. Specifically, we employ a tungsten lamp as a broadband light source, whose emitted light is focused onto the sample through a series of lenses. The focused light interacts with and transmits through the sample, producing mixed transmission signals containing different wavelengths and wavevectors. It is needed to point out that, as we are dealing with a linear system, light components with different wavelengths and wavevectors couple with the sample independently, making post-coupling resolution feasible. The transmitted light was then directed through a series of convex lenses which resolves different wavevector components by imaging the back focal plane - this plane contains the momentum-space (Fourier-space) information of the sample's radiation field. The wavevector-resolved transmission light was directed to the entrance slit of an imaging spectrometer (Princeton Instruments IsoPlane-320) for wavelength resolution. To isolate the resonant response from system artifacts, we normalize all measurements by dividing them by a background transmission spectrum obtained through an unetched, uniform Si layer. The transmission spectrum presented in the main text represent these normalized results, enabling clearer resolution of the band structure features.

**Iso-frequency contour and scattering spectrum measurement system:** A near-infrared laser with continuous tunability serves as the incident light source in this setup. The laser beam undergoes spatial filtering and is adjusted to match the sample size. Employing a polarizer, the laser polarization is tuned to align with the excitation k-point of the iso-frequency contours in the sample (Pol y). Following this, lens L1 focuses the light onto the back focal plane of an infinitely corrected objective. By displacing L1 in the x-y plane, the incident angle of the laser is controllable, allowing for oblique incidence excitation at any angle within the numerical aperture range. After traversing the sample, the scattered light is gathered by another objective and ultimately directed onto a CCD through successive lenses L2, L3, and a 4f system. The CCD corresponds to the Fourier plane of the sample. In the receiving optical path, an additional polarizer, oriented orthogonally to the incident polarization direction (Pol x), is utilized to suppress direct transmission through the sample, thereby enhancing the signal-to-noise ratio of the scattered light.

When the wavelength and incident angle of the incident light match the resonant modes supported by the sample, iso-frequency contours can be observed on the CCD. At this point, with the position of lens L1 fixed (i.e., without changing the incident angle), adjusting the laser can obtain the scattering signals corresponding to different wavelengths. When the wavelength matches the resonant mode, the iso-frequency contour is illuminated. As the wavelength gradually deviates from the resonance wavelength, the iso-frequency contour gradually dims. Thus, we can obtain the scattering spectrum curve of the intensity at any point in momentum space as a function of wavelength, and the Q-factor of which can be obtained by numerically fitting the scattering spectrum with a Lorentzian function.

**Numerical simulations:** All numerical simulations were conducted using the finite element method with commercial software COMSOL Multiphysics. To ensure the accuracy of the numerically calculated results, the mesh size used in the FEM simulation is small enough to guarantee the convergence of the results, which means a higher mesh density leads to nearly the same results. Due to the continuous translational symmetry along the *z*-axis, our simulations can be reduced to a 2D model within the *xy*-plane. Floquet periodic boundary condition was applied along the *x*-axis, while perfectly matched layers were implemented on the top and bottom boundaries of the simulation domain. To calculate the amplitude distribution of Fourier coefficients for flat-band eigenfields in momentum space, $u(k_x)$, we first perform an eigenfrequency calculation of the PhC slab. This allows us to obtain the Bloch modes $\mathbf{E}(\tilde{k}_x, x, y_0)$ in the near-field region. It is noted that $u(k_x)$ is proportional to the amplitude of the Bloch modes within the 1D PhC slabs. Thus, each eigenmode is normalized across the target band. Finally, the normalized Bloch modes are substituted into Eq. (1) to calculate $u(k_x)$ in Figure 2.

**Acknowledgements.**

This work was supported by National Natural Science Foundation of China (12234004 (X.Z.) and 12422411 (W.Z.)), National Key Basic Research Program (2022YFA1404900 (X.Z.)) and Young Elite Scientists Sponsorship Program by CAST Grant (2023QNRC001 (W.Z.)). The authors also acknowledge support from the Synergic Extreme Condition User Facility (SECUF), China.